\documentclass[11pt,preprint]{aastex}

\slugcomment{Accepted to the Astrophysical Journal Supplement Series}

\newcommand{\GPB}{\mbox{\em{GP-B}}}
\newcommand{\IMP}{\mbox{IM~Peg}}

\newcommand{\masyr}{\mbox{mas~yr$^{-1}$}}

\newcommand{\mua}{\mbox{$\mu_{\alpha}$}}
\newcommand{\mud}{\mbox{$\mu_{\delta}$}}

\newcommand{\ptwo}{\mbox{Paper~II}}
\newcommand{\pthree}{\mbox{Paper~III}}

\newcommand{\pfive}{\mbox{Paper~V}}
\newcommand{\psix}{\mbox{Paper~VI}}
\newcommand{\pseven}{\mbox{Paper VII}}

\defcitealias{GPB-I}{Paper~I}
\defcitealias{GPB-II}{Paper~II}
\defcitealias{GPB-III}{Paper~III}
\defcitealias{GPB-IV}{Paper~IV}
\defcitealias{GPB-V}{Paper~V}
\defcitealias{GPB-VI}{Paper~VI}
\defcitealias{GPB-VII}{Paper VII}

\newcommand{\Ra}[4]{\mbox{${#1}^{\rm h} \; {#2}^{\rm m} \; {#3}\fs{#4} $}}
\newcommand{\dec}[4]{\mbox{${#1}\arcdeg \; {#2}\arcmin \; {#3}\farcs{#4} $}}

\shorttitle{Overview of VLBI for {\em Gravity Probe~B}}
\shortauthors{Shapiro et al.}

\begin{document}
 
\title{VLBI for {\em Gravity Probe B}. I. Overview}

\author{I. I. Shapiro\altaffilmark{1},
N. Bartel\altaffilmark{2}, M. F. Bietenholz\altaffilmark{2,3},
D. E. Lebach\altaffilmark{1},
J.-F. Lestrade\altaffilmark{4}, 
R.~R. Ransom\altaffilmark{2,5}, and
M. I. Ratner\altaffilmark{1}}

\altaffiltext{1}{Harvard-Smithsonian Center for Astrophysics, 60
Garden Street, Cambridge, MA 02138, USA}

\altaffiltext{2}{Department of Physics and Astronomy, York University,
4700 Keele Street, Toronto, ON M3J 1P3, Canada}

\altaffiltext{3}{Now also at Hartebeesthoek Radio Astronomy Observatory,
PO Box 443, Krugersdorp 1740, South Africa}

\altaffiltext{4}{Observatoire de Paris, Centre national de la
recherche scientifique, 77 Av.\ Denfert Rochereau, 75014 Paris,
France}

\altaffiltext{5}{Now at Okanagan College, 583 Duncan Avenue West,
  Penticton, B.C., V2A 2K8, Canada and also at the National Research
  Council of Canada, Herzberg Institute of Astrophysics, Dominion
  Radio Astrophysical Observatory, P.O. Box 248, Penticton, B.C., V2A
  6K3, Canada}

\begin{abstract} 

We describe the NASA/Stanford gyroscope relativity mission, {\em
Gravity Probe~B} (\GPB), and provide an overview of the following
series of six astrometric and astrophysical papers that report on our
radio observations and analyses made in support of this mission.  The
main goal of this 8.5 year program of differential VLBI astrometry was
to determine the proper motion of the guide star of the \GPB\ mission,
the RS CVn binary IM~Pegasi (\IMP; HR~8703).  This proper motion is
determined with respect to compact, extragalactic reference sources.
The results are: $-20.833$ $\pm $ 0.090 \masyr\ and $-27.267$ $\pm $
0.095 \masyr\ for, respectively, the right ascension and declination,
in local Cartesian
coordinates, of IM~Peg's proper motion, and 10.370 $\pm$ 0.074 mas
(i.e., 96.43 $\pm$ 0.69~pc) for its parallax (and distance).  Each
quoted uncertainty is meant to represent an $\sim$70\% confidence
interval that includes the estimated contribution from systematic
error.  These results are accurate enough not to discernibly degrade
the \GPB\ estimates of its gyroscopes' relativistic precessions: the
frame-dragging and geodetic effects.
\end{abstract}


\section{Introduction}
\label{sintro}

According to Einstein's theory of general relativity (GR), space-time
is affected by both the Earth's mass and its angular momentum.  The
kinematics of bodies orbiting the Earth are thereby altered from their
expected behavior based on Newton's theory of gravity.  In particular,
according to GR, the spin axis of an ideal, freely-falling gyroscope
near the Earth should exhibit two distinct non-Newtonian precessions
due to these two properties of the Earth.  These precessions can be
considered as rotations of any near-Earth inertial frame with respect
to the distant univere.  The NASA/Stanford {\em Gravity Probe~B}
(\GPB) satellite was
placed in low-Earth orbit on 2004 April~20 to measure these rotations.

The \GPB\ spacecraft provided a nearly freely-falling (``drag-free"),
magnetically shielded, and thermally stable environment for its set of
four close-to-identical gyroscopes (hereafter, ``gyros") of novel
design and unprecedented stability and accuracy.  The design and
performance of these gyros are documented at length elsewhere
\citep[see, e,g.,][and references therein]{Conklin+2008, Keiser+2009}.
We describe here the key features of this experiment, emphasizing the
dependencies on astronomical measurements.

The four gyros provided a fourfold redundancy to increase the
reliability of the experimental results.  Each gyro rotor consists of
a 3.8~cm-diameter quartz sphere with about a 40-atom-thick niobium
coating which is superconducting at temperatures below 1.8~K\@.  The
four gyros were separately electrostatically suspended within a
structure largely made from a single block of quartz, which provided a
rigid framework with respect to which the orientation of the spin axis
of each gyro could be measured.  In turn, the orientation of this
structure with respect to the distant universe could be determined as
a result of its rigid attachment to a guide telescope that was
``locked'' on a specific bright star, the ``guide star,'' when gyro
measurements were being made.  To maintain the rigidity of the
telescope (and its attachment to the gyro housing), the telescope
body, too, was made of solid quartz.  The placement of this entire
package within a large dewar containing, at launch, $\sim$2,400 liters
of superfluid helium allowed for $\sim$17~months of continuous
cryogenic operation of the gyros in orbit.  To fit within a dewar,
this telescope was not only limited in size to a 15~cm diameter
aperture, but also reduced in light-gathering power by $\sim$25\% due
to its required placement behind a stack of four vacuum windows built
into the neck of the helium dewar.

To reduce errors resulting from nonrelativistic torques on the gyro
rotors due to gravitational, electrical, and magnetic interactions
between the rotors and the \GPB\ spacecraft, all four gyros were
placed on the optical axis of the telescope and all four spin axes
were aligned with that axis to within 10 milliarcseconds (mas).  In
this configuration, a high degree of reduction of the time-averaged
nonrelativistic torques was achieved by slowly rolling the entire
spacecraft about this common axis throughout the mission; the roll
period was 77.5 seconds.  To further reduce systematic errors, the
spin-axis direction of the second and fourth gyros in this linear
array was the same, but their spin vectors were oriented $180\arcdeg$
opposed to those of the first and third gyros. The primary
astronomically relevant consequences of this spacecraft design were
that a single guide star had to be used for the entire \GPB\
experiment and that only a bright star could serve this function.

The orbit selected and subsequently achieved for the mission was also
largely dictated by the desire to separate the two relativistic
effects from each other and to minimize the vector average over the
mission of the nonrelativistic, gravitational torques on the gyro
rotors.  Numerical studies completed many years before launch led to
the choice of a nearly circular, $\sim$640~km altitude, polar orbit,
with the orbital plane to be oriented so as to ensure that the line of
sight to the eventually-selected guide star would lie, on average for
the mission, within $\sim$1.5 arcminutes of the orbital plane of the
spacecraft (G.~Keiser 2009, priv.\ comm.).
A polar orbit ensured that the relativistic precession of the spin
axes about the orbit pole, due to the orbital motion of the spacecraft
(usually termed the ``geodetic'' precession), would lead to the gyro
spin axes drifting in the north-south direction.  (There is also a
relatively very small geodetic contribution from the Sun, resulting in
a drift in ecliptic longitude, which is taken into account in the
analysis.)  On the other hand, the predicted ``frame-dragging
effect,'' due to the rotation of the Earth (often called the
``Lense-Thirring'' or ``gravitomagnetic'' effect), would lead to the
gyro spin axes drifting eastward or westward, depending upon the
instantaneous latitude of the gyros.
The key required results from the \GPB\ spacecraft data analysis are
therefore the north-south and east-west components of the orbit-averaged rates
of precession of the gyros with respect to the apparent direction of the
guide star.  Similarly, the key required astronomical inputs to the relativity
tests are the two components of the mean rate of the apparent angular motion of
the guide star with respect to the distant universe.  (Here and hereafter we
use the word ``mean" to denote an average for the time interval for which 
precession data were collected.  Also, we use ``apparent" to reflect
the fact that effects such as aberration are not  completely averaged out over the mission.)
Of additional importance is the distance to the guide star, needed to make the
annual parallax correction in the analysis of the \GPB\ data; our VLBI program 
provides this value with high accuracy (see below).  

The time periods over which the gyros were monitored to collect
precession data were confined to the interval from 2004 August~27 to
2005 August~15.  The guide star was visible from the spacecraft for
about half of each orbit.  Generally, the attitude control system of
the spacecraft could lock the orientation of the spacecraft to the
direction of the guide star within ca.\ 2 minutes of its coming into
view.  Only data obtained when the spacecraft was locked on the guide
star were used to estimate the relativistic precessions of the gyros.

When averaged over an integer number of contiguous complete orbits,
each relativistic precession can be considered to occur at a uniform
rate.  For the orbit of \GPB, general relativity implies \citep[see,
e.g.,][]{Keiser+2009} that the average rate of geodetic precession is
$\sim$6\farcs{6}~yr$^{-1}$, while that due to frame-dragging is
$\sim$39~\masyr. The nominal, pre-launch, accuracy goal of \GPB\ was a
$\le$0.5~\masyr\ standard error in each precession.  At the time of
launch, the mission error budget allowed for a 0.15~\masyr\
contribution from astronomical phenomena to each of these two standard
errors; by far the most important contributor to each was the
allowance for the proper motion of the guide star.  The two largest
contributors to the overall error budget in this pre-launch analysis
were the residual nonrelativistic torques on the gyros and the noise
in the readout of the gyros; for discussion, see \citet{Keiser+2009}.
 
The above description of the \GPB\ mission implies that several
astronomical considerations and measurements were crucial to its
success.  A very distant extragalactic object would provide a
nearly ideal
tie to the distant universe.  But all such objects are far too dim to
be used directly.  A bright intermediary was therefore required: the
guide star.  A prime consideration was the selection of that star.
The obvious requirements were that it be sufficiently bright and
isolated on the sky, and also suitably located to nearly maximize the
sensitivity of the frame-dragging test.  Another critical requirement
for the guide star was that its proper motion either already be known
at the required accuracy or be measurable to that accuracy.  In 1989,
when \GPB\ seemed poised to enter mission status, it was clear, as it
had been previously, that there was no bright star whose proper motion
was known to even close to 0.15~\masyr.  The {\em Hipparcos}
astrometry spacecraft \citep{Perryman+1992,PerrymanH1993}, launched in
that year, was not expected to achieve such accuracy.  On the other
hand, many years earlier very-long-baseline interferometry (VLBI) at
centimeter wavelengths had yielded submilliarcsecond relative position
accuracy for compact extragalactic radio sources nearby to one another
on the sky \citep[see,
e.g.,][]{Shapiro+1979,MarcaideS1983,Bartel+1986}, and hence was
capable of yielding the desired accuracy in the proper motion of a
guide star that was visible at radio wavelengths (I.S., ca.\ 1975,
priv.\ comm.\ to C.~W.~F.\ Everitt).  Moreover, by 1990
submilliarcsecond accuracy had also been obtained for a faint radio
emitting star with respect to an extragalactic radio source
$\sim$$1\arcdeg$ away on the sky \citep{Lestrade+1990}.  
Therefore,
the \GPB\ project gave increasing attention to the ultimately adopted
option of using VLBI to determine the proper motion of an optically
bright radio star.  A combination of spacecraft engineering
requirements and the results of our program of radio observations of
various guide-star candidates led to the selection in 1997 of
IM~Pegasi (\IMP; HR~8703) \objectname[HR 8703] as the guide star.  We
describe the investigations that led to the selection of this
chromospherically active binary star in \S\ \ref{sselect} of this
paper.

Once \IMP\ had been selected, the bulk of the astronomical effort went
into the determination of its proper motion via a sustained VLBI
observation program, primarily at 8.4~GHz ($\lambda \simeq
3.6$~cm). We also utilized earlier VLBI observations of \IMP\ at this
same radio frequency made from 1991-94 by \citet{Lestrade+1999}.  In
addition, at our behest, many groups made observations of different
kinds at frequencies from ultraviolet to radio.  The major motivation
for these latter observations was the need to measure or bound any
difference between the proper motion of \IMP\ as determined with the
VLBI technique and the proper motion of \IMP\ as it would be observed
by the \GPB\ spacecraft in the wavelength range 0.3 to 1.1~nm.
Although this difference was never expected to be so large as to
degrade the accuracy of the mission, several possible contributions to
this difference had to be investigated observationally to meet the
mission requirement for an exceptionally high level of confidence in
its tests of general relativity.  These contributions are highly
dependent on the properties of \IMP, which we discuss in \S\
\ref{sprop} of this paper.

There were two other ways in which the \GPB\ experimenters made use of
astronomical knowledge.  First, because the \GPB\ spacecraft
continually rotated with a period of 77.5~s about the line of sight to
the guide star \citep[to within 0\farcs{2} rms when
locked;][]{Keiser+2009}, determination of the north-south and the
east-west components of the relativistic precession of the gyros
required knowledge of the roll phase of the spacecraft.  This roll
phase was needed to transform both the gyro orientation measurements
and the guide-telescope-pointing measurements from the frame of the
rotating spacecraft to a quasi-inertial frame.  The roll phase was
modeled based upon the outputs of the CCD detectors of two small
star-tracking telescopes that were fixed to the outside of the
spacecraft, aimed at angles of 50$\arcdeg$ and 60$\arcdeg$ to the
optical axis of the main guide telescope.  With each roll of the
spacecraft, each star tracker viewed an 8$\arcdeg$ wide band of the
sky.  The inference of spacecraft roll phase from the CCD readouts
required adequate knowledge of the astrometric positions of the
brighter stars in these bands.  In fact, the $\sim$$1\arcsec$ accuracy
requirement on these positions was easily satisfied by existing
astronomical catalogs
and required no new observations.  

Second, the aberration of the light from the guide star had to be
determined to compute the orientation of the spacecraft from the \GPB\
telescope readout data.  Thus the constantly changing spacecraft
velocity with respect to the solar-system barycenter (SSB) had to be
computed.  Spacecraft tracking data, as well as data from an on-board
GPS receiver, were used to determine the spacecraft velocity with
respect to the center of the Earth; the velocity of the Earth with
respect to the SSB was calculated from planetary and lunar
ephemerides.  Similarly, the relative positions of the Earth, Sun, and
planets were needed to compute the guide star's apparent motion due to
parallax and the deflection of the star's light by the Sun's mass.
Existing ephemerides exceeded by several orders of magnitude the
accuracies of velocities and positions needed for the \GPB\ mission.
The Galactocentric acceleration of the SSB can be neglected \citep{SoversFJ1998},
and presumably also the effect of a putative Nemesis solar companion
or some as yet unknown nearby dark cloud \citep[see also \pthree\ of this series,][]{GPB-III}.  
These latter possibilities will either be confirmed or bounded at a
useful level after, e.g., the Large Synoptic Survey Telescope goes
into operation.

Another critically important application for the \GPB\ mission of the
very accurately known values of the amplitudes, periods, and phases of
the aberration components of \IMP\ is their use as essentially
error-free calibrators for the conversion of the SQUID readouts of the
gyros from electrical to angular units \citep{Keiser+2009}.  All in
all, the design of the \GPB\ spacecraft assured that the astronomical
effects on its relativity mission could not only be adequately determined,
but also could be used to advantage in the analysis of the spacecraft
data.

This paper is the first of a series of seven describing the astronomical effort
we undertook to support the \GPB\ mission.  In the preceding paragraphs
we indicated the range of astronomical information required by the
mission.  Below we specify the requirements quantitatively, and outline
how they were met.  We then describe the six 
following papers in this series.  In addition, we
document certain aspects of the program that can be logically and adequately covered here.

In \S~\ref{sselect}, we describe the history behind the selection of
\IMP\ as the guide star, including the scope and results of an
$\sim$60~hr VLA search for radio emission from $\sim$1200 bright
stars.  We then summarize in \S~\ref{sprop} the stellar and orbital
properties of the \IMP\ binary, and comment on some significant
characteristics of its location in the sky, based on extensive
observations primarily at optical wavelengths.  Section \ref{sref}
contains descriptions of the compact, extragalactic radio sources used
to determine the guide star's proper motion 
with respect to the distant universe; 
\S~\ref{svlbiobs} mainly describes our procedures for
making VLBI observations of these sources.  In \S~\ref{ssynopsis}, we
summarize the six specialized papers of this series.  Section
\ref{sblind} treats our initial, but now dashed, hopes for a
``double-blind" experiment, and \S~\ref{sconc} lists our main
conclusions.

\section{Selection of IM PEG as the Guide Star}
\label{sselect} 

An obvious requirement for the guide star was that it be sufficiently
bright. As the spacecraft design evolved, this requirement became much
less stringent than had been envisioned earlier.  For many years, the
design called for photomultiplier tubes to be used as the light
detectors of the star-tracking system.  Not until the mid 1990's was
it considered safe to assume that the final design would, instead, use
photodiode detectors, which had higher quantum efficiency, peaking at
$\sim$80\% between 0.5~nm and 1.0~nm.  They also generated so little
heat that they could be placed inside the helium dewar without boiling
off helium at a rate that would significantly shorten
the cryogenic lifetime of the spacecraft, in fact $<1$\% (J.~Turneaure
2009, priv.\ comm.).  This internal placement, behind each of the two
focal planes of the telescope removed the need for light pipes to
transmit the starlight out of the dewar, and hence reduced the
transmission losses in the system.  Based upon consideration of the
resulting photon noise, B.~Lange (1994, priv.\ comm.)
estimated that a guide star could be marginally bright enough even at
V = 10.7 for stars of spectral type G or K\@.  However, the expected
levels of noise in the amplifiers used to generate the telescope
read-out signals implied that the maximum truly acceptable guide star
V magnitude was $\sim$7 (J.~Kasdin 1994, priv.\ comm.).  Both of these
limits were set with the intention that the uncertainty of the
measured pointing of the spacecraft averaged over operationally
relevant intervals
would not unacceptably degrade the real-time attitude control of the
spacecraft.  The minimum brightness required to avoid having telescope
read-out noise degrade the final estimates of the relativistic
precessions of the gyros was less stringent, and so automatically
satisfied \citep{Keiser+2009}.

For most of the period of development of the \GPB\ mission, the
nominal guide star was the very bright Rigel (V = 0.1).  In 1989, when
we began to intensively investigate the possibility of replacing Rigel
with a radio star observable with VLBI, one of our first tasks was to
investigate what the optically brightest suitable radio star might be.
Before describing our efforts and conclusions concerning that
question, we first specify what additional factors went into
evaluating the suitability of guide-star candidates.

First and foremost, a low declination for the guide star maximized the
sensitivity of the mission to the frame-dragging effect, since the
magnitude of that effect on the motion of the gyro spin axes on the
sky was proportional to the cosine of the spin-axis declination, which
had to be very nearly the same as that of the chosen guide star.
Since the errors in the measurements of gyro drift rates were not
expected to depend sensitively on declination, the fractional accuracy
with which the frame-dragging effect could be measured would be
maximized by a declination value of 0\arcdeg.  On the other hand,
if the expected error in the VLBI determination of the proper motion
of a given guide-star candidate were significantly smaller than that
for another candidate nearer to 0\arcdeg\ declination, the former
candidate might nevertheless have been a better choice than the
latter.  However, as was correctly anticipated, if the gyro drift-rate
measurements were considerably less accurate than the proper motions
determined by VLBI, then any compromise on the declination criterion
could significantly decrease the fractional accuracy of the
frame-dragging test.  Nevertheless, out of concern that we might find
no fully suitable radio star with the preferred low declination, we
examined stars with declinations as high as $+60\arcdeg$ and as low as
$-30\arcdeg$.  We considered any star yet further south to be too
difficult a target for accurate VLBI astrometry, since the antennas
available to us were predominantly located in the northern hemisphere.

Working against the declination preference was a strong preference for
a guide star more than $\sim$$20\arcdeg$ from the ecliptic.  For guide
stars at this or a greater angular distance from the ecliptic, a sun
shield could be placed around and in front of the telescope windows to
largely prevent sunlight from entering the dewar and boiling off the
helium even at the time of year when the sun was closest to the
direction of the guide star.  The scattering of direct sunlight by the
windows was also a potential source of error for the star tracker.  In
case this ecliptic-separation criterion was later relaxed, we included
in our guide-star search candidate stars that were as little as
$10\arcdeg$ from the ecliptic.

In 1996, as the time for selection of the guide star approached, it
became clear that stars much higher in absolute ecliptic latitude than
$30\arcdeg$ were unacceptable.  Because the spacecraft was to
continuously rotate about the line of sight to the guide star, while
its solar panels had to remain fixed on the spacecraft, the
roll-averaged amount of electrical power from them varied with the
ecliptic latitude and the time of year.  During the early and mid
1990's both the power requirements of the spacecraft and the expected
power output of the solar panels evolved.  Only in 1997 did the
project team conclude that, for the range of then practical spacecraft
designs, adequate power could not be guaranteed for stars as far as
$\sim$$40\arcdeg$ from the ecliptic.  As discussed below, this problem
directly impacted the final choice for the guide star.

The design of the guide-star telescope made it sensitive to light from
all astronomical sources within $\sim$$80\arcsec$ of the guide star:
With the spacecraft locked on the guide star to within a few
arcseconds, light from this surrounding field of view would fall on
the detectors.  Moreover, the smaller the ratio between the guide-star
brightness and that of background stars, the stricter the requirement
became for accurate knowledge of the time dependence during the
mission of that brightness ratio.  This restriction, too, as we'll see
below, resulted in the elimination of an otherwise promising
guide-star candidate.

The most fundamental requirement on the guide star is that its proper
motion be known, or known to be measurable, with sufficient accuracy
in an adequately inertial reference frame.
What was this accuracy requirement?  Given that the \GPB\ team wished
to perform the two relativity tests as accurately and convincingly as
feasible, and given that the tests each required an additive
correction for the proper motion, any specific requirement needed to
be justified in terms of the relative costs and benefits of reducing
the uncertainty contributed by each source of experimental error.
Only in 2003, the final year before launch, did the project approve a
formal requirement.  It called for the standard error in the estimate
of the proper motion of the guide star over the course of the mission
to be no more than 0.14~\masyr\ in each coordinate.  This somewhat odd
value was specified so that the total uncertainty due to all
astronomical phenomena would have a standard error no more than
0.15~\masyr\ after allowance was made for an additional standard error
of 0.05~\masyr\ for the independent effect of any background light in
the guide-star telescope's field of view.

The 0.15~\masyr\ requirement was chosen in light of highly uncertain
estimates of the non-astronomical experimental errors.  The nominal
goal of the mission design was to measure each relativistic effect
with a total standard error $\le$0.5~\masyr.  However, there was at
that time no identifiable reason why, in the event of a flawless
mission, the \GPB\ gyro measurements could not collectively yield a
full order of magnitude higher accuracy.  At the time (2003) that the
proper-motion requirement was formalized, we could predict with good
reliability that the series of VLBI observations we began in 1997, if
continued through the end of the \GPB\ flight mission, could meet the
0.14~\masyr\ requirement, even with the analysis allowing for the
possibility of a long-term proper acceleration due to an as yet
unknown, bound companion to the chosen guide star in an orbit with
period of several decades or more (see \S\ \ref{sprop}, below,
concerning shorter periods).  Higher accuracy in the proper-motion
determination could be obtained by continuing the VLBI measurements as
long into the future as required.

The above discussion of the constraints on the guide star is
summarized in Table \ref{tconstr}.  These constraints, however, are
not rigid, and the choices for guide star were quite limited, as we
discuss in the following paragraphs.
 
\begin{deluxetable}{l c}
\tabletypesize{\small}
\tablecaption{Constraints on Choice of Guide Star\label{tconstr}}
\tablewidth{0pt}
\tablehead{
Characteristic  & \colhead{Approximate constraint}  
}
\startdata
Brightness (mag) & $V \lesssim 7$ \\
Declination (deg) & $-20 \lesssim \delta\ \lesssim +20$ \\
Distance, $D$, from ecliptic  (deg) & $20 \lesssim D\lesssim 40$ \\
Minimum magnitude difference between guide star\\
~~~~and any background star within $4"$ -- $40"$ (with \\
~~~~gradual relaxation outside this range) (mag) & 10 \\
Standard error in final estimate of each component \\
~~~~of guide-star proper motion (\masyr) & $\le$ 0.14 \\
\enddata
\end{deluxetable}

Ground-based optical astrometry could not provide proper motions with
the required accuracy.  For example, in the Fifth Fundamental
Catalogue \citep[FK5,][]{Fricke+1988}, the mean individual error of
proper motion in right ascension for stars with $\delta > -30\arcdeg$
is $\sim$6~\masyr.  The {\em Hipparcos} satellite, launched in 1989,
unfortunately into the wrong orbit \citep{PerrymanH1993}, didn't seem
destined to be able to meet the \GPB\ requirements.  Even after the
miraculous completion of the {\em Hipparcos} program, resulting from
clever ``workarounds," the published catalog
\citep{PerrymanESAshort1997} shows that virtually all nominal standard
errors of the proper motions are greater than 0.5~\masyr\ in each
coordinate.  Moreover, in spite of a large, multi-pronged effort to
tie the reference system of the {\em Hipparcos} Catalogue to a
VLBI-governed International Celestial Reference Frame,
the uncertainty of the rate components of
that frame tie was at least 0.25~\masyr\ \citep{Kovalevsky+1997},
which was also expected to be unacceptable for \GPB.  In addition,
since the median epoch of the {\em Hipparcos} observations was
1991.25, a proper acceleration of the guide star, due to an undetected
bound stellar companion in a long-period orbit, plausibly could have
caused the apparent proper motion of that star during the year of the
\GPB\ mission to be subject to an offset larger than the nominal
standard error of the star's estimated proper motion at epoch 1991.25.
Worse is the later Tycho-2 Catalogue \citep{Hog+2000}, based on the combination of
{\em Hipparcos}/Tycho positions (but not the associated proper motions) 
and all usable ground-based positions (spanning
about a century). These provide proper motions with an estimated standard error 
of  2.5~\masyr\ in each coordinate.  
Finally, although modern optical methods can achieve differential positional accuracy over
single instrumental fields of view on the order of 1~mas in a single night, it
has not yet been demonstrated that a decade or so of such observations can yield
proper-motion standard errors even as low as 0.2~\masyr. 

In contrast to the apparently inadequate accuracy of these optically determined proper
motions, the accuracy of the upper bounds on the proper motion of compact, 
extragalactic radio sources derived from VLBI observations well exceeds the
requirements of \GPB.  

How do all of these considerations combine to affect the choice of the
guide star?  From our VLA survey (1990-1992) and among previously
known radio stars, we found only four potentially satisfactory guide
stars: $\lambda$~Andromeda \objectname[HR 8961]\ (HR~8961,
$+46\arcdeg$ declination), HR~1099 \objectname[HR 1099]\
($+1\arcdeg$), HR~5110 \objectname[HR 5110]\ ($+37\arcdeg$), and \IMP\
(HR~8703, $+17\arcdeg$).  All were known to be RS Canum
Venaticorum-type radio emitters before we conducted our radio survey
at 8.4~GHz of about 1200 other stars with V magnitude of 6.0 or
brighter.  Thus our survey yielded no detections of previously
undetected stellar radio emission.  Our results were confirmed by a
substantially deeper, more comprehensive, later survey by
\citet{Helfand+1999}, which disclosed no further stars suitable to be
GP-B guide stars; all failed on either or both of brightness and
declination grounds.
Following up on the four candidate guide stars, we examined the fields around each 
and checked as well on 
possible reference sources:  compact extragalactic radio sources nearby on the sky to each 
candidate.  
The \GPB\ project, in consultation with us, concluded that \IMP\ was the best choice;
the corresponding frame-dragging of the spacecraft was predicted to be $\sim$40~\masyr;
HR~5110 was a reasonably close second.  
Its elimination was based mainly on its high ecliptic latitude of $+43\arcdeg$.  

HR~1099 was rejected as the GP-B project was unsure whether, in the data analysis, 
the variation in the ratio of the brightness of HR~1099 ($V \sim$ 5.7) to that of a $V \sim$ 8.8 star
that would also be in the telescope's field of view could be measured to the accuracy needed
to avoid possibly degrading the accuracy of the relativity measurements.  
In addition, each time the telescope initiated its lock on the guide star, 
there would have been a risk of locking on the wrong star, at least temporarily.
The last remaining alternative to \IMP, 
$\lambda$~Andromedae,
was dropped because of its high declination and weak and variable  (typically 0.4 to 
$\sim$1 mJy) radio emission.   

\section{Properties of \IMP\ and its Surroundings}
\label{sprop} 

\IMP\ is a known binary star with a variable magnitude; see Table \ref{topt} for the observed 
maxima and minima of its V magnitudes during the \GPB\ mission.  The sky position of \IMP\ as 
well as its orbital elements are also shown in this table.  For  \IMP, 
no other star within $12'$ is 
brighter than V magnitude 10, compared with a corresponding average brightness of \IMP\ of about 6.
 
\begin{deluxetable}{l c c}
\tabletypesize{\small}
\tablecaption{Optical properties of \IMP\label{topt}}
\tablewidth{0pt}
\tablehead{
\colhead{Property}  & \colhead{Value}  
}
\startdata
{\em Hipparcos}\tablenotemark{a} 1991.25 RA (J2000) & \Ra{22}{53}{02}{278706}  $\pm$ 0.63~mas \\
{\em Hipparcos}\tablenotemark{a} 1991.25 DEC (J2000) &\dec{16}{50}{28}{53982}  $\pm$ 0.43~mas \\
Approximate galactic longitude, $l^{\rm{II}}$ & $\phantom{-}86.4\arcdeg$\\
Approximate galactic latitude, $b^{\rm{II}}$ &  $-37.5\arcdeg$\\
{\em Hipparcos} Parallax\tablenotemark{a} (mas) &  10.33 $\pm$ 0.76\\
{\em Hipparcos} Distance (pc) &  96.8 $\pm$ 7.1\\
Spectral type of primary\tablenotemark{b} & K2~III \\
V magnitude range:\tablenotemark{c} & 5.7 to 6.0\\
\sidehead{Spectroscopic orbital elements:\tablenotemark{d}}
Period, $P$ ( (days)  & 24.64877 $\pm$ 0.00003 \\
Eccentricity, $e$ & 0.0 \\
Superior conjunction of primary,\tablenotemark{e} $T_{\rm{conj}}$ & 2,450,342.905 $\pm$ 0.004 \\
Mass ratio, $M_2/M_1$ & 0.550 $\pm$ 0.001 \\
Velocity amplitude, $K$ (km s$^{-1}$) & 34.29 $\pm$  0.04 ~~~~~~~~~ 62.31 $\pm$ 0.06\\
$a$ sin $i$ (R$_\sun$) & 16.70 $\pm$ 0.02 ~~~~~~~~~ 30.34 $\pm$ 0.03 \\
Mass function, $f(m)$  (M$_\sun$) & ~~0.1030 $\pm$  0.0004 ~~~~~~ 0.618 $\pm$ 0.0002 \\
$M$~sin$^3$~$i$ (M$_\sun$) & ~~1.486 $\pm$  0.007 ~~~~~~~$\;$ 0.818 $\pm$ 0.005 \\
\\
Orbital inclination,\tablenotemark{f} $i$  & $ 65\arcdeg \le i \le 80\arcdeg$
\enddata
\tablenotetext{a}{ \citet{PerrymanESAshort1997}.  The position is given for the catalog epoch, 1991.25.}
\tablenotetext{b}{\citet{BerdyuginaIT1999} 
and \citet{Marsden+2005}.}
\tablenotetext{c}{ Near daily photometry, save for $\sim$2.5 months when the Sun
prevented observations, was obtained by G.~Henry (2005, priv.\ comm.)\ during the mission.}
\tablenotetext{d}{\citet{Marsden+2005}. The values for the two binary components are given in two columns.}
\tablenotetext{e}{ Julian date for heliocentric observations.
Note:  Superior conjunction refers to a body's being furthest from us in its orbit.}
\tablenotetext{f}{\citet{BerdyuginaIT1999}.  See also \citet{Lebach+1999}, who find 
$i \gtrsim 55\arcdeg$. }
\end{deluxetable}

We made quite extensive, but unpublished, investigations into possible 
systematic errors of the \GPB\ measurements due to both known and 
unknown, but plausible, optical properties of \IMP\ and the field of view 
of the \GPB\ guide telescope when it was locked on \IMP.   
We were particularly concerned that the photospheric spots 
(analogous to sun spots, but much larger) that characterize the primary component of 
RS CVn binaries like \IMP, could cause the apparent center of the guide-star telescope
image of \IMP\ to systematically drift with respect to our VLBI-derived proper motion of
the center of mass of the system.  A program of spot mapping by 
S.~Marsden  and S.~Berdyugina of ETH-Zurich \citep{Marsden+2007}, 
using optical spectroscopic observations, found no such effect, 
and ruled out errors larger than 0.04 \masyr.   Among the 
spectroscopic observations used to reach this conclusion were
two full observing seasons of near nightly observations, effectively covering the entire 
\GPB\ mission.  These data were obtained by J. Eaton of Tennessee State University
with the TSU Automated Spectroscopic Telescope (Fairborn Obs., Paradise Valley, AZ).

A second class of conceivable
errors encompassed all those that could arise due to the photometric variability 
of \IMP\ in combination with sensitivity of the \GPB\ 
telescope to ``background" light in its field of view; such latter effects could arise 
from point sources and nebulosity, whether 
constant or variable.  These possibilities, too, were ruled out, 
on the basis of a wide variety of observations obtained in support of \GPB.  
Notable among these observations were images obtained by us as we searched for 
unknown stellar companions (or nebulosity) 
with the WFPC2 instrument on HST (using filters ranging from IR to UV),
by P.~Kalas (UC Berkeley) with his ``coronagraphic" camera on the 
U.~Hawaii 2.2~m telescope (Mauna Kea),
by L.~Roberts (then at Boeing) with the Advanced Electro-Optical System Telescope
(USAF Res.\ Lab., Haleakala),
by E. Horch (U. Mass. Dartmouth) via speckle interferometric observations using
WIYN (Kitt Peak), and by X.~Pan (Caltech) with the Palomar Testbed Interferometer.
T.~Dame (CfA) used the CfA's 1.2 m aperture radio telescope to map 
the sky near \IMP\ in CO(1$-$0) millimetric 
emission to rule out any compact molecular cloud near \IMP\ that might 
be associated with an optical reflection nebula.
Based on these observations and a Bayesian probabilistic analysis 
by J.~Chandler (CfA) and one of us (MIR), any astrometric errors due to 
undetected companions of \IMP\ were bounded below 0.006 mas/yr 
with about 95\% confidence, under plausible but conservative 
assumptions about the a priori  distribution of third-body companions
with respect to optical brightness, orbital parameters, and other 
characteristics.  In addition, extensive optical photometry of \IMP\ 
by G.~Henry (TSU) using mainly the TSU Automatic Photometric Telescope
(Fairborn Obs., Paradise Valley, AZ), 
and of the known background stars in the guide-telescope field of view
by G. Gatewood (U. Pittsburgh), using the Allegheny Obs.\ Thaw telescope (Pittsburgh),
and by A. Henden (then USNO), using the 1.55 m USNO telescope (Flagstaff), 
bounded any photometric variation of those stars during the \GPB\ 
mission sufficiently to rule out any significant resulting error in the 
\GPB\ measurements.  All together, the above errors do not
contribute as much as the 0.05~\masyr\ standard error allowed 
for them in the \GPB\ error budget.

\section{Radio Reference Sources for IM~Peg}
\label{sref} 

\begin{figure}[tp]
\centering
\includegraphics[height=4in,trim=1in 4.6in 0.0in 1.4in,clip]{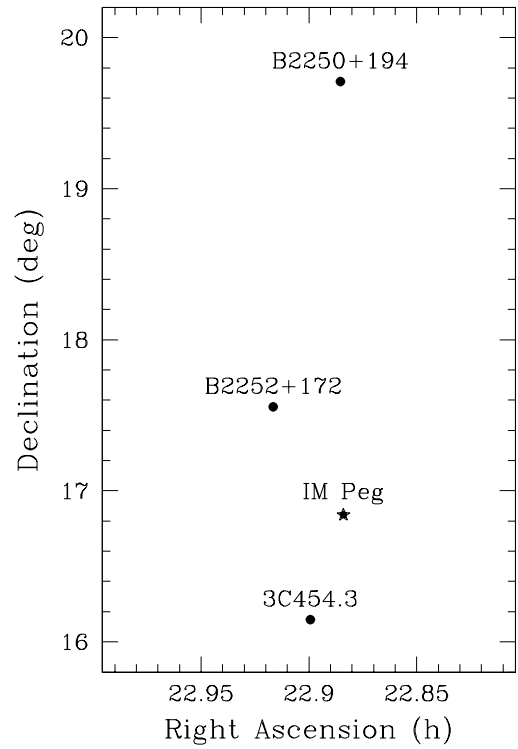}
\caption{Positions (J2000) on the sky of the four radio sources used for \GPB\ astrometry. 
The east-west and north-south directions on the plot are shown to the same scale.
}  
\label{fskypos} \end{figure}

Our main goal is to determine the proper motion of \IMP\ with respect to the distant universe.
To this end, we sought compact, extragalactic radio sources which were effectively 
fixed markers in the distant universe.  By using phase-referenced VLBI,
we could determine the difference with time between the positions of \IMP\ and those of the
chosen extragalactic radio sources.   
What compact extragalactic radio sources did we choose?  The main 
reference source we chose, 3C~454.3\objectname[3C454.3], is located on the sky 
$0.7\arcdeg$ away from \IMP.  This reference source has a complicated, changing radio 
brightness distribution.  
Nonetheless, we made this choice, also motivated by 3C~454.3 having been used as
the reference source for the VLBI observations of \IMP\ in 1991-1994 
\citep{Lestrade+1999}.  We wanted to thereby take advantage of the extended
time span that would then be available for our determination of proper motion
\citep[see \pfive,][for discussion of the value of the earlier observations]{GPB-V}.
To check on possible systematic errors that 
might affect the VLBI determinations of the sky position of \IMP\ 
with respect to 3C~454.3, we added two more reference sources.  One of these sources, the quasar 
B2250+194 (ICRF J225307.3+194234)\objectname[ICRF J225307.3+194234]\ was 
included ab initio, and was also used to distinguish between model errors that have
elevation-angle dependence and those that do not.  The other, 
B2252+172 (87GB 225231.0+171747)\objectname[87GB 225231.0+171747], 
was added in 2002; although quite close to \IMP\ and virtually a point source,
it is a weak radio emitter and not always reliably detected. 
These latter two sources are far more compact than 3C~454.3, but are further 
away on the sky from \IMP\ or very weak, as noted.  
The former is $2.2\arcdeg$ away and the latter $0.9\arcdeg$ away;
see Figure~\ref{fskypos}  for the relative sky positions of all four sources.  
The redshift of 3C~454.3 is 0.859, whereas that of B2250+194 is 0.28.  The third 
reference source does not have a known redshift, but its compact structure, flat microwave 
power spectrum \citep[see \ptwo,][]{GPB-II}, and lack of any detectable proper motion (see below) 
constitute in sum virtual proof of its extragalactic nature.   

\section{VLBI Observations for \GPB}
\label{svlbiobs}

The VLBI data that we gathered in each of our 35 observing sessions 
between 1997 and 2005 were obtained at 8.4-GHz, 
with the addition of 5 and 15~GHz data for one observing session
\citepalias[see][]{GPB-II}.  In each session, we used
up to 16 antennas distributed globally.  
Our choice of 8.4~GHz as the primary observing frequency was based on 
its yielding the best combination of high sensitivity and high angular resolution
when our full array of antennas was used.
For each such session, observations were made
in a repeating cycle that included the guide star and each reference source.   This cycle 
extended over 5.5~minutes or somewhat ($<$20\%) longer 
for the earlier sessions and consisted of interleaved cycles of 
duration 5.5 and 7 minutes for the last 12 sessions, when the third reference source was included.  
(The latter pattern was a compromise we adopted to more nearly preserve the SNRs for our first 
two reference sources.)  The cyclic pattern of observing was designed to
reduce the effects on our results of systematic errors due to our inability to adequately model 
the temporal behavior at each antenna site of the clocks and the propagation medium, the latter
consisting most importantly of the Earth's atmosphere and ionosphere.
(Had multiple antenna
beams been available at each site, simultaneous observations 
could have been made of all of the target sources, obviating the need for the cyclic observing.)  
The downside of this mode is the difficulty in properly
connecting the observed fringe phase from each cycle to the next 
so as to eliminate multiple 2$\pi$ ambiguities (see below).  Based on signal-to-noise-ratio 
considerations, we also chose to not observe simultaneously in two radio bands, 
even though such dual-band observations would have allowed us to largely free our VLBI
data from ionospheric effects.

\section{Synopsis of Series of Papers}
\label{ssynopsis}

This section is devoted to a synopsis of each of the remaining papers of this series.

\citetalias{GPB-II} focuses on mapping and analyzing the changing radio-brightness structures 
of the three compact extragalactic reference sources used in our determinations of the proper 
motion of \IMP, the \GPB\ guide star.  \citetalias{GPB-II} also describes our VLBI observations in detail,  
the reference sources used for each session, the processing
needed to produce maps of the brightness distributions of those sources 
and of the guide star, and the resultant reference-source maps themselves
(the maps of \IMP\ are given in \citetalias{GPB-VII}; see below).  Figure~\ref{fimages} 
shows a typical example of an 8.4 GHz image of the guide star and of each reference source.  
A summary of all of these observations---epochs, wavelengths, antennas, and sources---is presented
in \citetalias{GPB-II}.  Through the analysis of the radio images from our 35 sessions
(spread over 8.5 years) of VLBI observations of  3C~454.3, our principal reference source, this paper 
establishes that a specific brightness component, dubbed C1, at the eastern end of the source, 
likely corresponds to the gravitational center of the source (the ``core")
and to a putative supermassive black hole located there (see Figure~\ref{fimages}).  
Small, under 0.2~mas, excursions of the brightness peak of C1 from this 
core location were tracked from session to session.  These motions are plausibly attributed to the 
effects of occasional outbursts from the core which manifest themselves as jets, 
initially unresolved, that move the 
peak of C1 westward.  Then, as the jet separates further, the location of the peak becomes less 
affected and hence moves eastward back towards its ``normal" position collocated with the core.  
This interpretation of C1 is bolstered by C1's steep spectrum  and by comparison of our 8.4~GHz 
images with contemporaneous and near contemporaneous images at 43 and 86~GHz frequencies.  
In addition, \citetalias{GPB-II} 
shows maps and presents analyses of
the temporal evolution of the other compact components in the structure of 3C~454.3.
The paper also establishes the utility of our two other reference sources as relatively structureless, 
nearly unchanging secondary reference sources for our VLBI astrometry of  \IMP.

\begin{figure}[tp]
\centering
\includegraphics[height=2.25in,trim=0 0.0in 0 0.2in,clip]{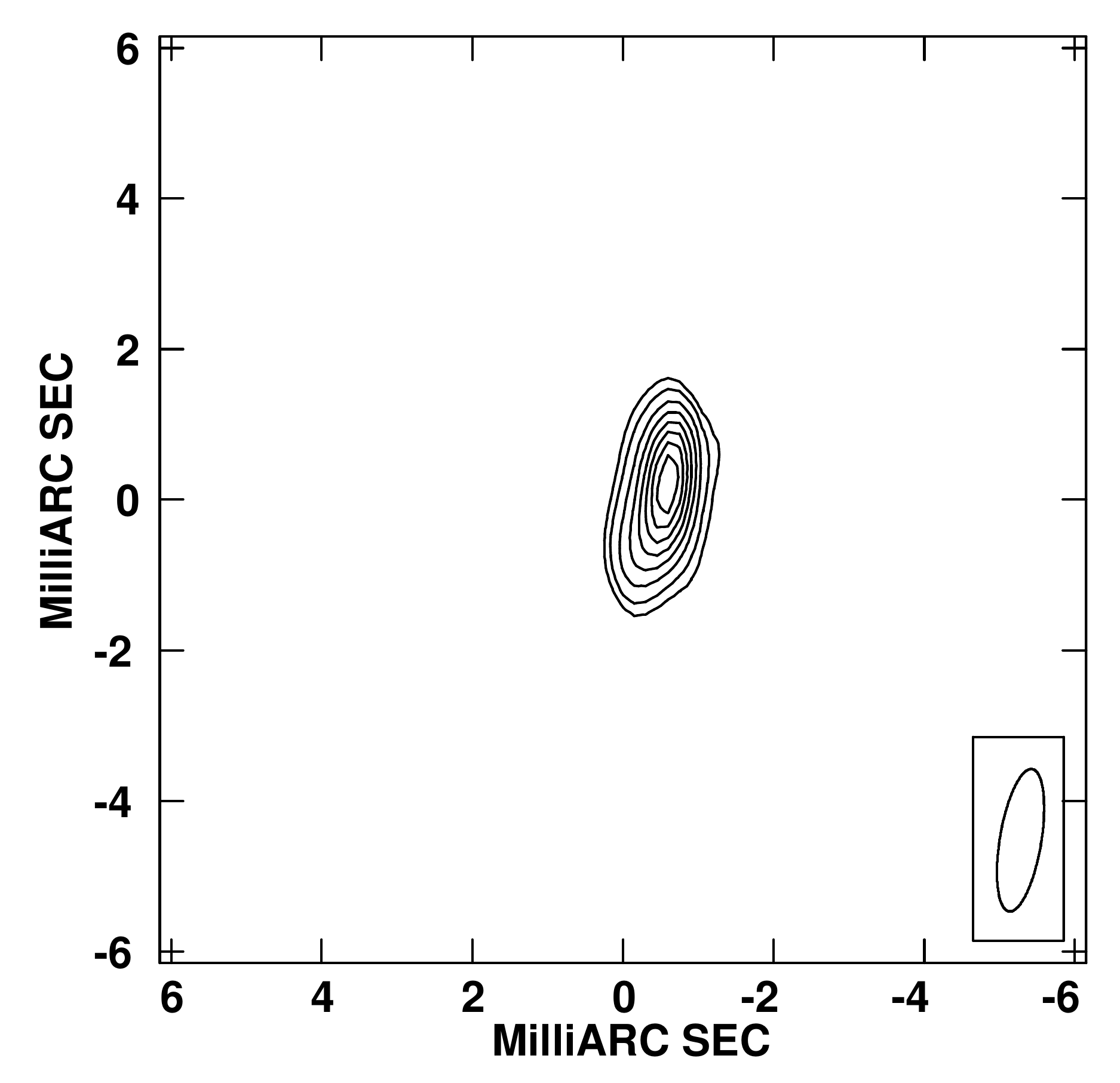}
\includegraphics[height=2.35in,trim=0 0.0in 0 0.2in,clip]{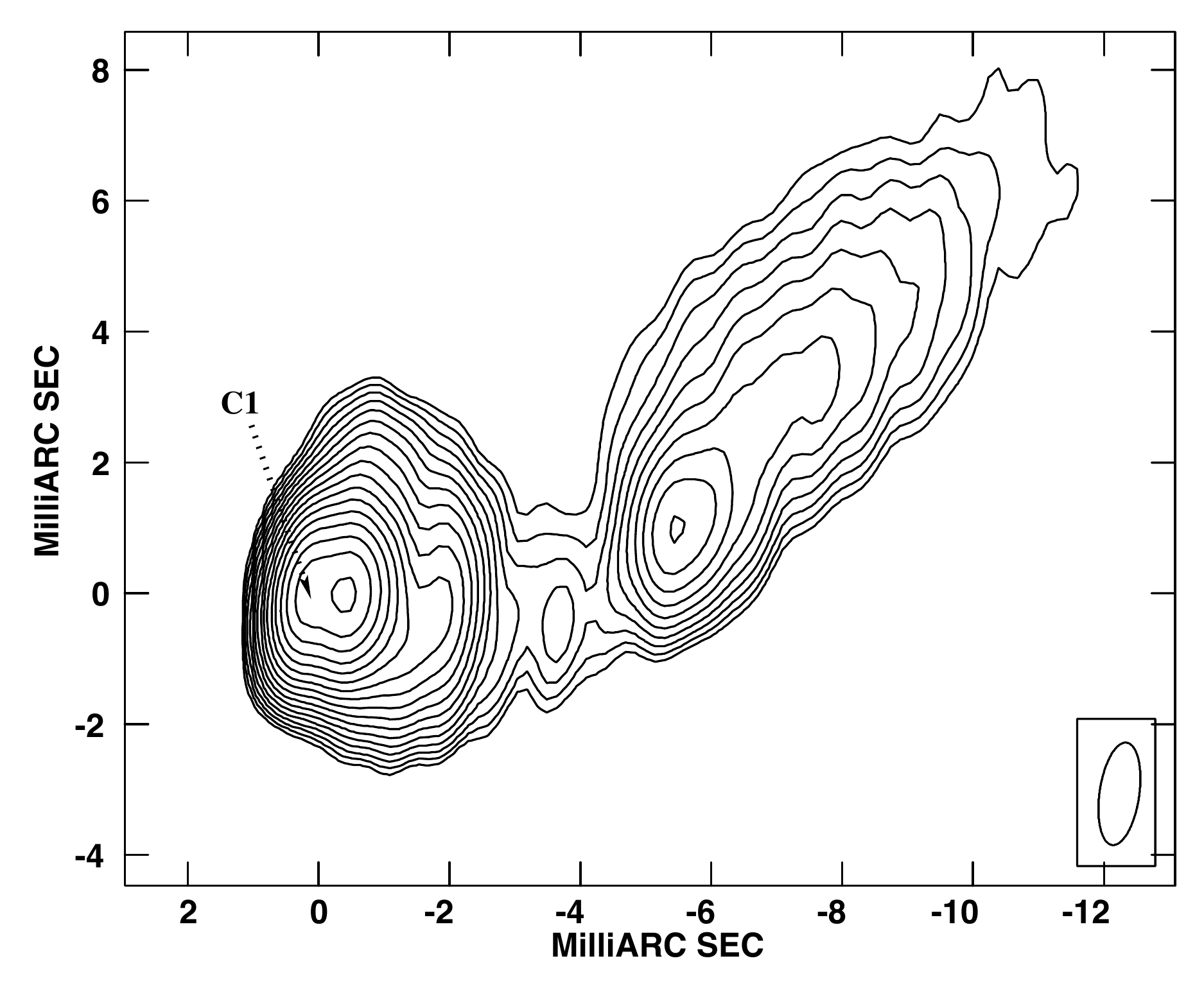}
\includegraphics[height=2.8in,trim=0 1.9in 0 1.0in,clip]{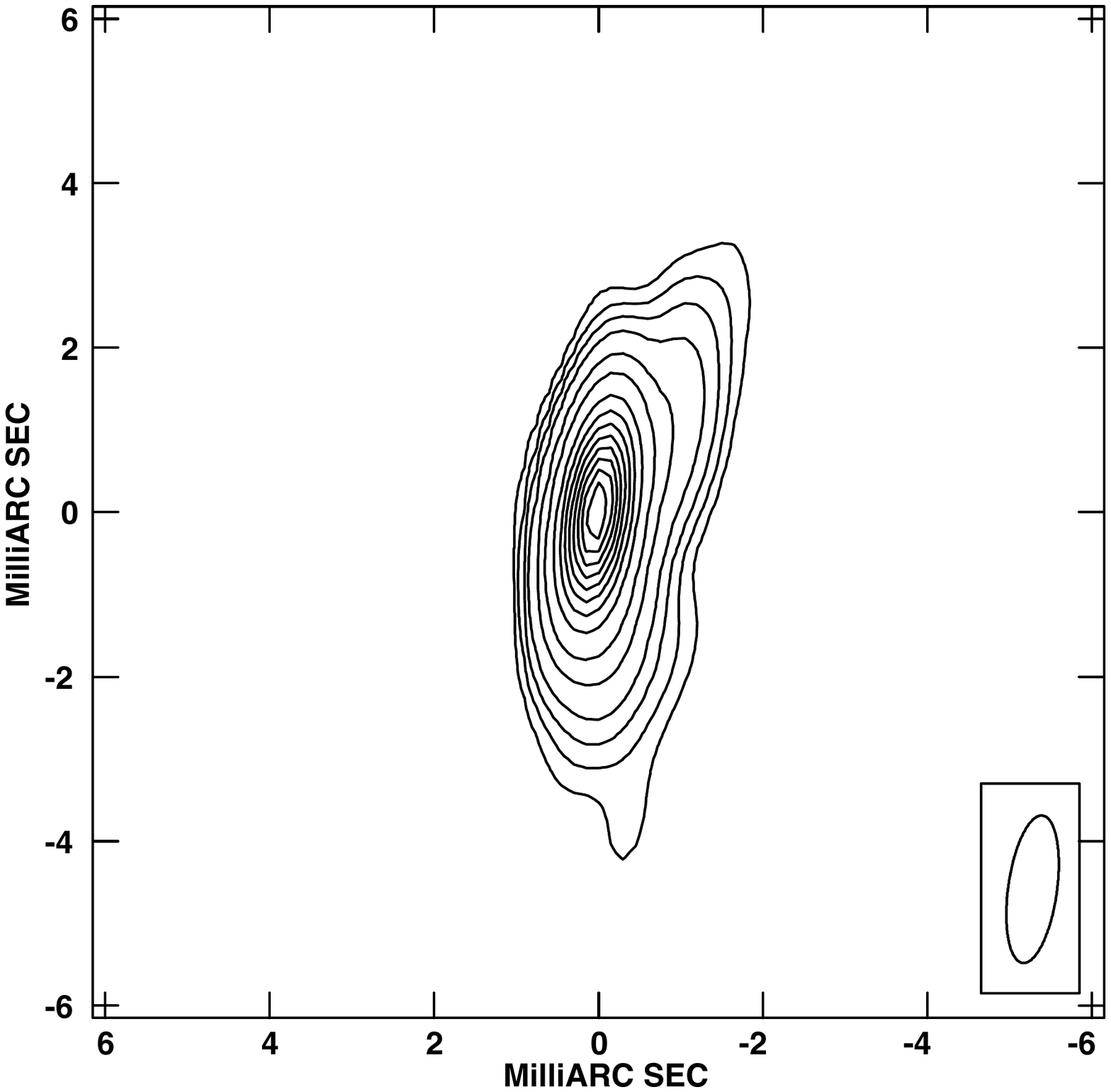}
\includegraphics[height=2.8in,trim=0 1.9in 0 1.0in,clip]{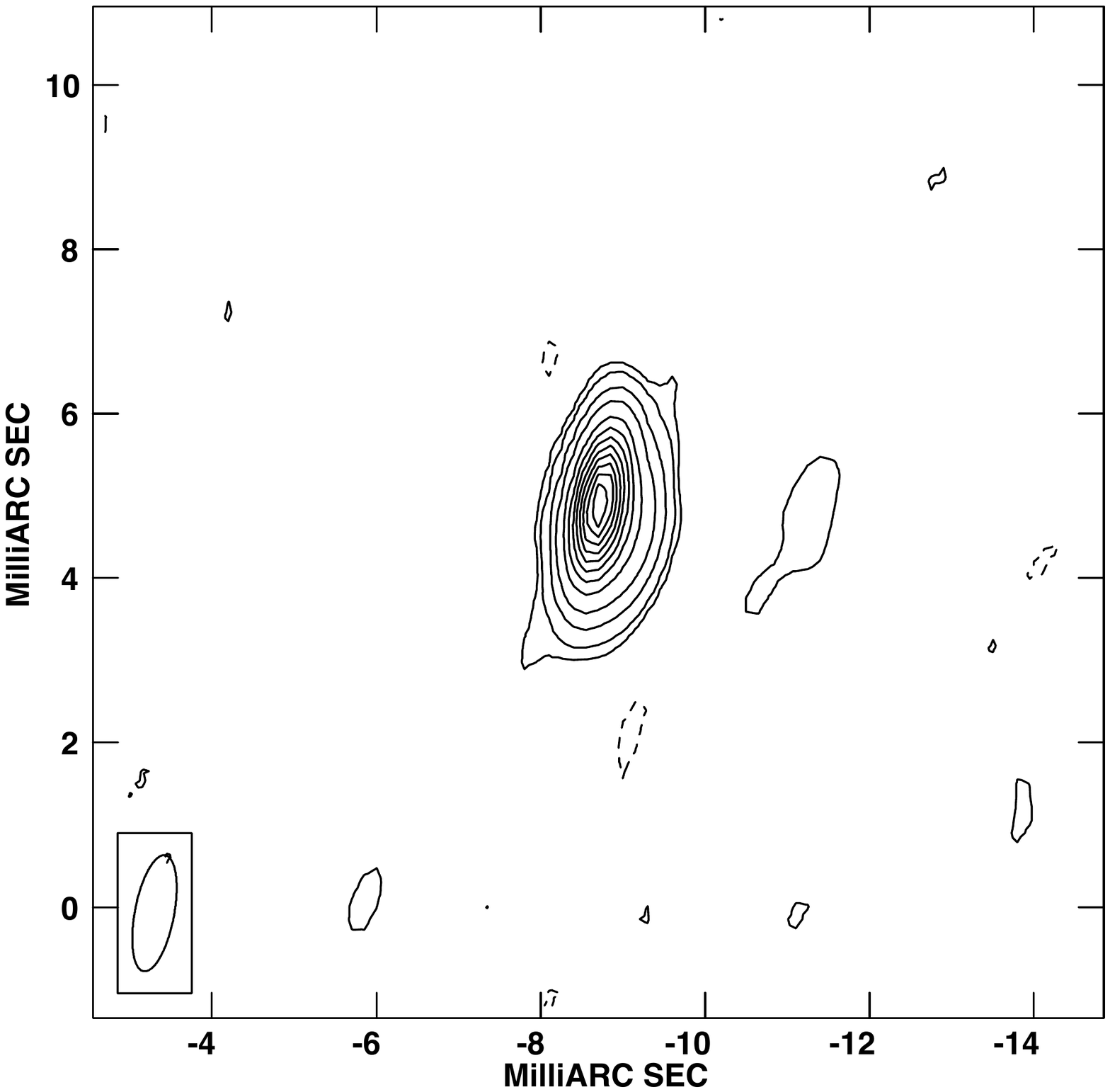}
\caption{
Sample CLEANed VLBI images at 8.4 GHz of \IMP\ and our three VLBI reference sources, all
at approximately the same scale.  North is up and east to the left.   The positions of the origins are 
not significant here.  The upper left panel shows one of our higher SNR images of \IMP, derived from 
all usable data from our 2004 December~11 observing session.
As at most of our observing epochs, the detectable stellar radio emission consists of a single slightly 
extended component with little or no visible structure.  The upper right panel shows 3C~454.3 at the 
same epoch.  A well resolved core-jet structure is seen in all of our 3C~454.3 images. 
The component labeled C1, which we found to have no significant astrometric motion with respect to 
the peaks of our two other extragalactic sources,  serves as our astrometric fiducial point (see text).  
The lower left panel shows  B2250+194 on 2000 November~05.  Extensions to the northwest and south are evident.
The last panel (lower right) shows B2252+172, our most compact reference source, on 2003 May~8.  
In each panel the restoring beam used in processing the image is shown in the inset.
These images are presented in the relevant papers of our series; see these for contour details.}  
\label{fimages} \end{figure}

\pthree\  \citep{GPB-III} delves deeply into the structure and behavior of the radio
brightness of our main reference source, 3C~454.3.  The primary goal of this examination for \GPB\ 
was to determine a stable feature in its radio brightness distribution, one which remained at a fixed 
location with respect to the center of mass of the source.  
This study led to our choice of C1 (see Figure~\ref{fimages}) as the
reference position in this source.  From our full set of maps of
3C~454.3, one for each observing session, we were able to follow the
evolution of this source's radio brightness at 8.4~GHz with reasonably
good time resolution over nearly a decade.  A main thrust of
\citetalias{GPB-III} is to establish, to a degree of reliability
sufficient for the \GPB\ mission, that the C1 component of 3C~454.3 is
stationary with respect to the distant universe, approximated by
positions of extragalactic reference sources.  We established this
stationarity in \citetalias{GPB-III}, primarily by i) using our VLBI
phase-delay observations to determine the position of C1 with respect
to the positions of our other two radio reference sources, and then
ii) determining the position of one of the latter (B2250+194) with
respect to those of a large suite of compact extragalactic radio
sources.  To this latter end, we made use of the extensive $\sim$30
years' accumulation of astrometric/geodetic group-delay VLBI
observations of 3C~454.3 \citep[see][and references
  therein]{Petrov+2009} that determine its position in a catalog
formed from such observations of $\sim$4000 of these sources that were
observed rather regularly over this period.

\citetalias{GPB-III} concludes that $\sim$70\% confidence upper limits
on the proper motion of C1 on the plane of the sky for the time period from 1998 to 2005 are
0.046 \masyr\ in the right-ascension and 0.056 \masyr\ in the declination
directions.
These limits notwithstanding, \citetalias{GPB-III} also presents
evidence for C1 having a ``jittery" east-west motion,
with amplitude $\sim$0.2 mas, 
likely related to jet activity in the vicinity of the core, as
discussed above for \citetalias{GPB-II}.  \citetalias{GPB-III} also
analyzes in detail the proper motions of the other radio-bright
components of 3C~454.3, some superluminal.

In \citetalias{GPB-IV}
we describe the novel 
data-reduction technique we used for \GPB\@ in our effort 
to achieve as high an astrometric accuracy from our VLBI data as feasible.
Our technique combines the superior model-correction capabilities of
parametric model fits to VLBI data
with the ability of phase-referenced maps to yield astrometric
measurements of sources that are too weak to be used in parametric
model fits.  More specifically, we use VLBI data from our radio-bright
reference sources in parametric model fits to improve our 
a~priori models (in particular for propagation delays through the
Earth's atmosphere), and then use these improved models to make
phase-referenced maps of our target sources. 
As shown in \citetalias{GPB-IV}, this technique has
benefits for both our astrometry and the dynamic range of our
target-source maps.  Our technique also allowed astrometric results
with submilliarcsecond accuracy to be obtained from each of our 35
sessions of VLBI observations of \IMP, an outcome that may not have
been possible with conventional techniques that use parametric model
fits or phase-referenced maps alone.  \citetalias{GPB-IV} also
describes our successful strategy for removing 2$\pi$ ambiguities from
the fringe-phase data from the observations of our reference sources,
a key element of our data-reduction technique.

\pfive\ in the series \citep{GPB-V} contains our astrometric analysis
of the time series of positions for the radio source associated with
\IMP, determined as described in \citetalias{GPB-IV}.  Using a
weighted-least-squares algorithm, we determined the parameters, their
uncertainties, and their correlations, for a model (and several
variants) of the motion of this radio source.  Each of these models
has parameters representing sky position at a reference epoch,
parallax, proper motion, and the orbit of the close binary system.
The orbit is assumed circular with a known period, based on optical
spectroscopic observations \citep[see, e.g.,][]{Marsden+2005}; we also
found consistency between our data and a zero eccentricity orbit, as
well as with the optically-derived orbital period, which is determined
far more accurately from the far longer series of optical
spectroscopic data.  The main alternative model considers the possible
presence of a distant third body in the guide star's system.  This
presence would lead, over the short term, to a proper acceleration as
well as a contribution to the proper motion of the \IMP\ binary.
However, the estimates of the associated parameters yield values
insignificantly different from zero.  Our final result for \IMP's
proper motion and parallax (see Table~\ref{tfinal}) is thus based on a
9-parameter model: 4 orbital parameters, 2 for sky position (right
ascension and declination of the center of mass of the \IMP\ binary at
epoch), 2 components of its proper motion, and 1 parallax parameter
(see Table 3 of \citetalias{GPB-V}).

\begin{deluxetable}{l@{\hspace{80pt}}c@{\hspace{70pt}}c}
\tabletypesize{\scriptsize}
\tablecaption{\IMP\ parameter estimates \label{tfinal}}
\tablewidth{0pt}
\tablehead{
Parameter   & \colhead{Estimate}  
& \colhead{Total SE\tablenotemark{a}}  }
\startdata
$\mua$ (\masyr)\tablenotemark{b} &  $-20.833$  
& 0.090 \\
$\mud$ (\masyr) &  $-27.267$ 
& 0.095 \\
Parallax (mas) &  \phn 10.370   & 0.074 \\
Distance (pc) & \phn 96.43 & 0.69\phn \\
\enddata
\tablenotetext{a}{Each ``Total SE" entry is our estimate of the
parameter's standard error, inclusive of both statistical and
estimated systematic errors, as described in \citetalias{GPB-V}.}
\tablenotetext{b}{The notation here denotes the $\alpha$
component of proper motion multiplied by $\cos{\delta}$,
i.e., the local Cartesian coordinate in the right-ascension
direction.}

\end{deluxetable}

\psix\ \citep{GPB-VI} examines the orbit of the radio emission of
\IMP, as projected on the sky, and compares the common properties with
those deduced from optical spectroscopy.  From the projected orbit and
the different radio images ---with one, two, or three
components---obtained at different epochs, Ransom et al.\ develop a
simple model of the radio emission.  Simulations based on this model
point to the brightness peaks of the radio emission at the various
epochs of observation emanating preferentially from over the polar
rather than over the equatorial regions.  The sky-projected mean
position of these peaks lies within about 35\% of the distance from
the center to the surface of the primary.  Another inference is that
about two-thirds of the peaks originate at altitudes below about 25\%
of the radius of the primary.

\pseven\ \citep{GPB-VII} focuses on the images of the guide star for
all of our observing sessions.  The image for \IMP\ for each session
was made via the phase-referenced mapping method.  Included in the
paper is a short movie that shows the temporal behavior of the radio
brightness distribution of the guide star over our $\sim$8.5~years of
VLBI observations.  Unfortunately, the sparse and uneven spacing of
the epochs of observation make the presentation somewhat ``choppy."
But two main points 
are clear: (1) the time-variable sky positions of the radio source,
relative to the putative position of the primary component of the
guide-star binary system, and (2) the  time-variable brightness
distribution of the source.
Each of these appears to change chaotically with time, albeit
within the ``reasonable" ranges also discussed in Papers V and VI.  In
an attempt to explain some of the aspects of the
brightness-distribution changes, the paper proposes a speculative
model based on an assumed dipolar magnetic field of the primary.  This
model finds some support in the comparison of its predictions with the
observed positions and shapes of \IMP's radio brightness distribution
at our observing epochs.

\section{Double-Blind Experiments}
\label{sblind}
				
It is best that an experimenter not know the ``right" result in
advance.  Why?  To avoid possible bias, whether conscious or
unconscious.  This possibility of bias is probably most important to
suppress in experimental tests of GR.  Present experiments are far
from probing any quantum limit of GR and the a priori
expectation among physicists is extremely high that far from this or
the so-called strong-field limit, this theory will be valid for
currently accessible accuracy levels.  In a typical experiment, those
involved carefully examine all its aspects to assess the likely level
of systematic errors as well as the contribution of random errors.
Although similar in principle, a GR experiment may be different in
practice: In essence, if experimenters obtain the ``wrong" answer,
they re-examine the experiment in excruciating detail, looking for a
possible error; when they obtain the ``right" answer, they publish.
To eliminate this type of bias, a double-blind experiment would be
ideal.  Although this approach is not often feasible in a physics
experiment, the \GPB\ mission, in principle, offered this opportunity.
The measurement of the changes in direction of the gyroscopes were
made with respect to the guide star, whose proper motion was indeed
already known but with an uncertainty about tenfold higher than the
expected uncertainty of the measurements of the gyro precessions.

One idea was to have the \GPB\ group at Stanford determine, without
our knowledge, the apparent proper motion of the guide star relative
to the \GPB\ gyros on the assumption that GR was correct, thus
approximating the proper motion of the guide star as we measured it,
i.e., with respect to the distant universe.
Our group, without the Stanford group's knowledge or involvement,
would determine the proper motion of the guide star with respect to a
nearly inertial frame defined by compact extragalactic radio sources
representing the distant universe.  After both groups had completed
their analyses, they would get together.  Our group's result would be
subtracted from Stanford's (or vice versa) and the result
checked to see whether it is zero to within the estimated
errors---confirmation of GR---or significantly different from
zero---incompatible with GR.  This comparison would take place in the
presence of knowledgeable neutral observers and possibly
representatives of the media -- an unusual scientific gathering!  Of
course, in the end, the double blindness of the experiment would
depend on the integrity of the members of each group to keep their
results totally to themselves until the comparison event.

Alas this plan was not to be carried out.  There were two problems:
\begin{itemize}	
\item[1.] The accuracy of the pre-mission value of the guide star's proper motion 
\citep{PerrymanESAshort1997} was improved 
about twofold and made public \citep{vanLeeuwenF2008}; and
\item[2.]  The accuracy of the \GPB\ measurements of the guide star's motion in the gyro frame 
decreased about twentyfold  compared with the pre-mission expectations.  Thus, the uncertainty 
published for the guide star's 
proper motion \citep[$\sigma\approx 0.3$ \masyr\ in each coordinate;][]{vanLeeuwenF2008}  
was substantially under that of
the \GPB\ determination of the motion 
\citep[$\sigma\approx$ 7 \masyr\ in right ascension and 18 \masyr\ in declination;][]{Everitt+2011}.
Within these larger-than-expected uncertainties, the three 
proper-motion estimates agreed.
\end{itemize}

Our VLBI determination  \citepalias{GPB-V} of \IMP's proper motion
thus becomes useful primarily as an accurate check, with $\sigma \le 0.10$ \masyr\ in each coordinate.

\section{Conclusion}
\label{sconc}

Our VLBI observations represent the most comprehensive set of radio 
images ever obtained on a star.  We find that:
\begin{itemize}
\item[1.] The proper motion of  \IMP\ on the plane of the
sky (i.e., in local Cartesian coordinates) is $-20.83$  $\pm$ 0.09 \masyr\ and
$-27.27$ $\pm$  0.09 \masyr\ in right ascension and declination, respectively;
\item[2.] The parallax and distance to \IMP\ are, respectively, 
10.37 $\pm$ 0.07~mas and  96.43 $\pm$ 0.69~pc; 
\item[3.] The centers of the maps obtained from our 35 sessions of
VLBI observations moved erratically from session to session with
respect to our estimate of the sky position of the primary component
of the binary guide star;
\item[4.] For one session there was a remarkable correlation
\citep{Lebach+1999} between rapid changes in total radio brightness of
\IMP\ and corresponding changes in sky position of the radio source,
at about the quarter-hour limits of our useful time resolution.  Other
sessions showed similar, but not as definite, relations between
changes in radio-source flux density and changes in its sky position.
These features of this radio star cry out for more quantitative
theoretical understanding than is provided by our mostly qualitative
speculations in \citet{Lebach+1999} and in \citetalias{GPB-VII}; and
\item[5.]  The $1\sigma$ uncertainty in our determination of the proper motion of
\IMP\ is about 30\% less than the accuracy goal of 0.15~\masyr\ 
set by the GP-B project.
\end{itemize}
 
 \acknowledgements 
ACKNOWLEDGEMENTS.  We would like to thank each of the following
people, listed alphabetically, for their contributions of observations
and data reductions to our investigations of \IMP\ and its sky
neighborhood, for providing us with unpublished updated catalog data,
and/or for enlightening discussions on various aspects of our research
program:
D.~Buzasi, 
A.~Buffington, 
N.~Caldwell, 
R.~Donahue, 
G.~Fazio, 
D.~Fischer, 
O.~Franz, 
E.~Guinan, 
P.~Hemenway, 
M.~Holman, 
the late J.~Huchra, 
G.~Keiser, 
J.~Kolodziejczak, 
R.~Kurucz,  
B.~Lange, 
D.~Latham, 
J.~Lipa, 
H.~McAlister, 
D.~Monet, 
R.~Schild, 
G.~Torres, 
W.~Traub, 
D.~Trilling, %
N.~Turner, 
the late H.~Wendker, 
and the late C.~Worley, 

This research was primarily supported by NASA, through a contract from
Stanford University to SAO, with a major subcontract from SAO to York
University.  We obtained observations made with the 100-m telescope of
the MPIfR (Max-Planck-Institut f\"{u}r Radioastronomie) at Effelsberg,
with antennas of the National Radio Astronomy Observatory (NRAO),
which is a facility of the National Science Foundation operated under
cooperative agreement by Associated Universities, Inc., with antennas
of the DSN, operated by JPL/Caltech, under contract with NASA, and
with the NASA/ESA Hubble Space Telescope, which is operated by the
Association of Universities for Research in Astronomy, Inc., under
NASA contract NAS 5-26555. These last observations are associated with
program \#07209.  We made use of the SAO/NASA Astrophysics Data System
Abstract Service, initiated, developed, and maintained at SAO.

\end{document}